\documentclass[seceq,preprint]{ptptex}

\preprintnumber[3cm]{
arXiv:0808.0355\\
RIKEN-TH-135\\ August, 2008}

\title{
 Comments on Gauge Invariant Overlaps for Marginal Solutions
 in Open String Field Theory
}

\author{
Isao \textsc{Kishimoto}\footnote{
E-mail:~ikishimo@riken.jp}
}

\inst{
Theoretical Physics Laboratory, RIKEN,
 Wako 351-0198, Japan
}

\abst{
We calculate the gauge invariant overlaps for
Schnabl/Kiermaier-Okawa-Rastelli-Zwiebach's marginal solution
with nonsingular current.
The obtained formula is the same as that for
Fuchs-Kroyter-Potting/Kiermaier-Okawa's marginal solution,
which was already computed by Ellwood.
Our result is consistent with the expectation that these
two solutions may be gauge equivalent.
We also comment on a gauge invariant overlap
for rolling tachyon solutions
in cubic open string field theory.
}

\begin{document}

\maketitle


\section{Introduction}

Since Schnabl's construction of an analytic solution 
for tachyon condensation\cite{Schnabl_tach}
in bosonic cubic open string field theory,
there have been new developments.
(See Ref.~\citen{Fuchs:2008cc} and references therein.)
In Refs.~\citen{Ellwood} and \citen{KKT1},
gauge invariants ${\cal O}_V(\Psi)$ 
specified by on-shell closed string states,
 which we will call gauge invariant overlaps,
were computed for some solutions.
The evaluation of gauge invariant overlaps,
in addition to that of the action,
can be used to check the gauge equivalence
of apparently different string fields.
In Ref.~\citen{KKT1}, the values of gauge invariant
overlaps for Schnabl's analytic solution\cite{Schnabl_tach}
and the numerical solution in the Siegel gauge 
(Ref.~\citen{Gaiotto:2002wy} and references therein)
for tachyon condensation were computed and compared.
The result was consistent with the expectation of
their gauge equivalence.
In Ref.~\citen{Ellwood}, gauge invariant overlaps for Schnabl's 
solution \cite{Schnabl_tach} and 
one type of marginal solutions given in 
Refs.~\citen{Fuchs:2007yy} and \citen{Kiermaier:2007vu}
were evaluated and interesting formulas were found.
In this work, we will compute gauge invariant overlaps for
another type of marginal solutions constructed in 
Refs.~\citen{Schnabl:2007az} and \citen{Kiermaier:2007ba}
 and find the same formula for marginal solutions as
obtained in Ref.~\citen{Ellwood}.
This is consistent with the expectation that
the marginal solutions given in
Refs.~\citen{Fuchs:2007yy} and \citen{Kiermaier:2007vu}
and in Refs.~\citen{Schnabl:2007az} and \citen{Kiermaier:2007ba}
are gauge equivalent.
We also apply our result to a rolling tachyon solution
investigated in Ref.~\citen{HS},
which is an example of the marginal solutions in 
Refs.~\citen{Schnabl:2007az} and \citen{Kiermaier:2007ba},
and comment on the large deformation limit.\\

Let us begin by reviewing marginal solutions and gauge 
invariant overlaps briefly.
There are two types of marginal solutions.\footnote{
Another type of marginal solutions, which is 
 based on the identity state, such as those in
 Refs.~\citen{TT1} and \citen{Kluson:2003xu}, is known.
In this case, the evaluation of gauge invariant overlaps is difficult
because of divergence due to the contraction of two identity states.
}
One was constructed by Schnabl\cite{Schnabl:2007az} and 
Kiermaier-Okawa-Rastelli-Zwiebach \cite{Kiermaier:2007ba},
which we abbreviate as Schnabl/KORZ's marginal 
solution.
The other one was constructed by Fuchs-Kroyter-Potting \cite{Fuchs:2007yy}
and generalized by Kiermaier-Okawa\cite{Kiermaier:2007vu},
which we denote as FKP/KO's marginal solution in the following.
The former can be applied only to the case of nonsingular 
marginal current $J$, namely, the operator product expansion (OPE)
among $J$s is nonsingular: $J(y)J(z)\sim {\rm finite}$ ($y\to z$).
For the latter, we can apply more general currents 
using a particular regularization, although we treat only nonsingular
current $J$ in this paper for simplicity.
Both solutions have one parameter $\lambda_{\rm m}$
and the same form, $\lambda_{\rm m}cJ(0)|0\rangle$,
for the lowest term with respect to $\lambda_{\rm m}$,
but the higher terms are different.

Schnabl/KORZ's marginal solution $\Psi_{\lambda_{\rm m}}^{\rm S/KORZ}$
is given by\footnote{
We use the notation in Ref.~\citen{Schnabl_tach}. 
$\Psi_{\lambda_{\rm m}}^{\rm S/KORZ}$
 is essentially obtained from a BRST-invariant and 
nilpotent string field
$\hat\psi_{\rm m}=\hat{U}_1\tilde c\tilde J(0)|0\rangle$
as\cite{Kmi}
$\Psi^{(\alpha,\beta)}_{\lambda_{\rm m}}=P_{\alpha}*(1+
\lambda_{\rm m}\hat\psi_{\rm m}*A^{(\alpha+\beta)})^{-1}*
\lambda_{\rm m}\hat\psi_{\rm m}*P_{\beta}
$,
where $P_{\alpha}=\hat{U}_{\alpha+1}|0\rangle$, 
$A^{(\gamma)}=\frac{\pi}{2}\int_0^{\gamma}d\alpha
B_1^LP_{\alpha}$, and $B_1^L=\frac{1}{2}(b_1+b_{-1})+\frac{1}{\pi}({\cal
B}_0+{\cal B}_0^{\dagger})$.
In this paper, we set $\alpha=\beta=1/2$ for simplicity.
Other solutions, except for $\alpha=\beta=0$, can be obtained using 
the relation 
$\Psi^{(\alpha,\beta)}_{\lambda_{\rm
m}}=e^{\frac{\pi}{4}(\beta-\alpha)K_1}
(\alpha+\beta)^{({\cal L}_0-{\cal L}_0^{\dagger})/2}
\Psi^{(1/2,1/2)}_{\lambda_{\rm
m}}$, which gives the same value of 
 gauge invariant overlaps thanks to Eq.~(\ref{eq:symm_Kn}).
}
\begin{eqnarray}
 \Psi_{\lambda_{\rm m}}^{{\rm S/KORZ}}
&=&\sum_{n=1}^{\infty}\lambda_{\rm m}^n\psi_{{\rm m},n}\,,\\
\psi_{{\rm
 m},k+1}
&=&\left(-\frac{\pi}{2}\right)^k\int_0^1 dr_1
\cdots\int_0^1 dr_k\,
\hat U_{\gamma^{(k)}+1}
\prod_{m=0}^k\tilde J(\tilde x_m^{(k)})\nonumber\\
&&\quad\times \biggl[-\frac{1}{\pi}({\cal B}_0+{\cal B}_0^{\dagger})
\tilde c(\tilde x_0^{(k)})\tilde c(\tilde x_k^{(k)})
+\frac{1}{2}\left(\tilde c(\tilde x_0^{(k)})+
\tilde c(\tilde x_k^{(k)})\right)
\biggr]|0\rangle\,,
\label{eq:psi_mk+1}
\end{eqnarray}
where $U_r\equiv (2/r)^{{\cal L}_0},\hat{U}_r\equiv U_r^{\dagger}U_r$,
${\cal B}_0\equiv
 b_0+\sum_{k=1}^{\infty}\frac{2(-1)^{k+1}}{4k^2-1}b_{2k}$
and ${\cal L}_0=\{Q_{\rm B},{\cal B}_0\}$.
The arguments of fields $\tilde c$ and $\tilde J$,
where $\tilde \phi(\tilde z)$ in the sliver frame is given by
$(\cos\tilde z)^{-2h}\phi(\tan \tilde z)$ using $\phi(z)$
in the upper half-plane for a primary field with dimension $h$,
are specified by
\begin{eqnarray}
&&\gamma^{(k)}=1+\sum_{l=1}^kr_l\,,~~~
\tilde x_m^{(k)}=\frac{\pi}{4}\biggl(\gamma^{(k)}-1-2\sum_{l=1}^m
r_l\biggr).
\end{eqnarray}
On the other hand, FKP/KO's solution is given by (Appendix
\ref{sec:FKP-KO})
\begin{eqnarray}
\Psi^{{\rm FKP/KO}}_{\lambda_{\rm m},L}
&=&\sum_{n=1}^{\infty}\lambda_{\rm m}^n\psi_{L,n}\,,
\label{eq:FKP/KOv1}
\\
\psi_{L,k+1}&=&
\hat{U}_{k+2}\tilde c\tilde J\Bigl(\frac{\pi}{4}k\Bigr)
\nonumber\\
&&\times (-1)^k
\int^{\frac{\pi}{4}k}_{
\frac{\pi}{4}(k-2)}\!d\tilde x_1
\int^{\tilde x_1}_{\frac{\pi}{4}(k-4)}\!d\tilde x_2
\cdots
\int^{\tilde x_{k-1}}_{\frac{\pi}{4}
(-k)}\!\!d\tilde x_k
\tilde J(\tilde x_1)\tilde J(\tilde x_2)
\cdots\tilde J(\tilde x_k)|0\rangle.
\nonumber\\
\end{eqnarray}
In order to satisfy the reality condition, we should apply a gauge
transformation by
\begin{eqnarray}
 {\cal U}&=&{\cal
  I}+\sum_{n=1}^{\infty}\lambda_{\rm m}^n{\cal U}_n\,,
\\
{\cal U}_n
&=&\hat{U}_{n+1}
(-1)^n\int_{\frac{\pi}{4}(-(n-1)
}^{\frac{\pi}{4}(n-1)}
d\tilde x_1
\int_{\tilde x_1}^{\frac{\pi}{4}(n-1)}
d\tilde x_2\cdots
\int_{\tilde x_{n-1}}^{\frac{\pi}{4}(n-1)}
d\tilde x_n\tilde J(\tilde x_1)\tilde J(\tilde x_2)\cdots 
\tilde J(\tilde x_n)|0\rangle,
\nonumber\\
\end{eqnarray}
where ${\cal I}=\hat{U}_1|0\rangle$ is the identity state, then
\begin{eqnarray}
 \Psi_{\lambda_{\rm m}}^{\cal U}
&\equiv&\frac{1}{\sqrt{{\cal U}}}*\Psi^{{\rm
  FKP/KO}}_{\lambda_{\rm m},L}
*\sqrt{{\cal U}}+\frac{1}{\sqrt{{\cal U}}}*
Q_{\rm B}\sqrt{{\cal U}}\,.
\label{eq:gauge_tr_U}
\end{eqnarray}

The gauge invariant overlap ${\cal O}_V(\Psi)$ is defined by
\begin{eqnarray}
 {\cal O}_V(\Psi)&=&\langle {\cal I}|V(i)|\Psi\rangle
=
\langle \hat{\gamma}(1_{\rm c},2)|V_{\rm c}\rangle_{1_{\rm
c}}|\Psi\rangle_2\,,
\label{eq:GIO_defn}
\end{eqnarray}
where $\langle {\cal I}|V(i)
=\langle \hat{\gamma}(1_{\rm c},2)|V_{\rm c}\rangle_{1_{\rm
c}}$ corresponds to 
an on-shell closed string state and
$\langle \hat{\gamma}(1_{\rm c},2)|$
is Shapiro-Thorn's vertex\cite{ST}, 
which relates the closed string Hilbert space $(1_{\rm c})$
to the open string Hilbert space $(2)$.
$|V_{\rm c}\rangle=c_1\bar c_{1}V_{\rm m}(0,0)|0\rangle$ is given by 
a matter primary field $V_{\rm m}(z,\bar z)$ with dimension $(1,1)$.
We consider $\Psi_{\lambda_{\rm m},L}^{\rm FKP/KO}$ instead of 
$\Psi_{\lambda_{\rm m}}^{\cal U}$
because gauge invariant overlap ${\cal O}_V(\Psi)$ is
invariant under gauge transformations.
In particular, the on-shell closed string state
in the open string Hilbert space,
$\langle {\cal I}|V(i)
=\langle \hat{\gamma}(1_{\rm c},2)|V_{\rm c}\rangle_{1_{\rm
c}}$, 
has the following symmetries (Appendix \ref{sec:S-T}):
\begin{eqnarray}
&&\langle {\cal I}|V(i)K_n=0\,,~~~~~~~
~~~~~~~K_n\equiv L_n-(-1)^nL_{-n}\,,
\label{eq:symm_Kn}\\
&&\langle {\cal I}|V(i)(b_n-(-1)^nb_{-n})=0\,,
\label{eq:symm_bn}\\
&&\langle {\cal I}|V(i)(c_n+(-1)^nc_{-n})=0\,.
\label{eq:symm_cn}
\end{eqnarray}
In the first line, $L_n$ denotes the total Virasoro generator,
which has zero central charge.
The first line can be derived from the second line and BRST invariance:
$\langle {\cal I}|V(i)Q_{\rm B}=0$.\\

The rest of this paper is organized as follows.
In the next section, we compute gauge invariant overlaps
for two types of marginal solutions by rewriting
string fields appropriately.
In \S \ref{sec:rolling}, we comment on gauge invariant overlaps
for lightlike and timelike rolling tachyon solutions
using the result in \S \ref{sec:evalation_m}.
In Appendix \ref{sec:Schn_tach},
 we rewrite Schnabl's solution for tachyon
condensation in the same way as marginal solutions 
in \S \ref{sec:evalation_m}.
In Appendix \ref{sec:FKP-KO}, we briefly review FKP/KO's marginal
solution and fix our conventions.
In Appendix \ref{sec:S-T}, we derive symmetries for 
on-shell closed string states 
using the Shapiro-Thorn vertex.

\section{Evaluation of gauge invariant overlaps for marginal solutions
\label{sec:evalation_m}
}

We will rewrite the marginal solutions
in order to evaluate gauge invariant overlaps easily
using symmetries (\ref{eq:symm_Kn}), (\ref{eq:symm_bn}) and
(\ref{eq:symm_cn}) of on-shell closed string states.

Schnabl/KORZ's marginal solution can be decomposed in the same way as
$\Psi_{\lambda}^{\rm S}$ (Appendix \ref{sec:Schn_tach})
because the ghost sector of $\psi_{{\rm m},k+1}$ (\ref{eq:psi_mk+1})
is similar to $\psi_r$ (\ref{eq:psirv1}). From
(\ref{eq:hatU_identity}), (\ref{eq:calB_identity})
and $r^{{\cal L}_0}\tilde \phi(\tilde z)r^{-{\cal L}_0}=r^h \tilde
\phi(r\tilde z)$ for a primary field $\phi$ with dimension $h$,
 we rewrite $\psi_{{\rm m},k+1}$ as
\begin{eqnarray}
 \psi_{{\rm
 m},k+1}
&=&\left(-\frac{\pi}{2}\right)^k\!\int_0^1\! dr_1
\cdots\!\int_0^1 \! dr_k\,
(\gamma^{(k)})^{({\cal L}_0-{\cal L}_0^{\dagger})/2}
\prod_{m=0}^k\!\left((\gamma^{(k)})^{-1}\tilde J\Bigl(\frac{\tilde
x_m^{(k)}}{\gamma^{(k)}}\Bigr)\right)
\nonumber\\
&&\times \Biggl[\frac{\gamma^{(k)}}{\pi}({\cal B}_0-{\cal B}_0^{\dagger})
\tilde c\Bigl(\frac{\tilde x_0^{(k)}}{\gamma^{(k)}}\Bigr)
\tilde c\Bigl(
\frac{\tilde x_k^{(k)}}{\gamma^{(k)}}\Bigr)
+\frac{1}{2}\Biggl(
\tilde c\Bigl(\frac{\tilde x_0^{(k)}}{\gamma^{(k)}}\Bigr)+
\tilde c\Bigl(\frac{\tilde x_k^{(k)}}{\gamma^{(k)}}\Bigr)\Biggr)
\Biggr]|0\rangle.
\end{eqnarray}
Then we have
\begin{eqnarray}
 \Psi_{\lambda_{\rm m}}^{\rm S/KORZ}
&=&\sum_{k=0}^{\infty}\lambda_{\rm m}^{k+1}\!\int_0^1\!\!dr_1\!
\cdots \!\int_0^1\!\!dr_k\!
\left(\frac{-\pi}{2}\right)^k\!(\gamma^{(k)})^{-k-1}\!
\prod_{m=0}^k\!\tilde J\Bigl(\frac{\tilde x_m^{(k)}}{\gamma^{(k)}}\Bigr)
c_1|0\rangle
\nonumber\\
&&+O({\cal L}_0-{\cal L}_0^{\dagger},{\cal B}_0-{\cal
 B}_0^{\dagger},c_n+(-1)^nc_{-n})\nonumber\\
&=&\sum_{k=0}^{\infty}\lambda_{\rm m}^{k+1}\!\int_0^1\!\!dr_1\!
\cdots \!\int_0^1\!\!dr_k\!
\left(\frac{-\pi}{2}\right)^k\!(\gamma^{(k)})^{-k-1}\!
\prod_{m=0}^k\!\tilde J\Bigl(\frac{\tilde x_m^{(k)}-\tilde
x_0^{(k)}}{\gamma^{(k)}}\Bigr)c_1|0\rangle
\nonumber\\
&&+O(K_1,{\cal L}_0-{\cal L}_0^{\dagger},{\cal B}_0-{\cal
 B}_0^{\dagger},c_n+(-1)^nc_{-n})
\,,
\label{eq:psiSKORZv2}
\end{eqnarray}
where we have used the relations: $
e^{\alpha K_1}\tilde J(\tilde z)e^{-\alpha K_1}=\tilde J(\tilde
z+\alpha)$,
(\ref{eq:ctil+ctil}), (\ref{eq:ctilctil})
and 
\begin{eqnarray}
&&e^{\alpha K_1}c_1|0\rangle=\tilde c(\alpha)|0\rangle=\tilde 
c_1|0\rangle
+\sum_{k=0}^{\infty}(\alpha^{2k+2}\tilde c_{-1-2k}|0\rangle
+\alpha^{2k+1}\tilde c_{-2k}|0\rangle),\\
&&\tilde c_{1-2k}|0\rangle=\sum_{l=0}^{k-1}{\cal C}_{l}^{(k)}
(c_{1-2k+2l}-c_{-1+2k-2l})|0\rangle,~~~~~~~(k=1,2,3,\cdots )\\
&&{\cal C}_{l}^{(k)}\equiv \oint_0\frac{dz}{2\pi i}(\arctan z)^{-2k-1}
(1+z^2)^{-2}z^{2k-2l}\!,~~~~
{\cal C}_{k-1}^{(k)}+{\cal C}_{k}^{(k)}=0,~~~{\cal C}_0^{(k)}=1,\\
&&\tilde c_1|0\rangle =c_1|0\rangle,~~~~~
\tilde c_0|0\rangle =c_0|0\rangle,\\
&&\tilde c_{-2k}|0\rangle=\left(\sum_{l=0}^{k-1}
{\cal C}_l'^{(k)}(c_{2l-2k}+c_{2k-2l})
+{\cal C}_k'^{(k)}c_0
\right)|0\rangle\,,~~~~(k=1,2,3,\cdots)\\
&&{\cal C}_l'^{(k)}\equiv \oint_0\frac{dz}{2\pi i}(\arctan z)^{-2k-2}
(1+z^2)^{-2}z^{2k-2l+1}\!,~~{\cal
C}_k'^{(k)}=\frac{(-1)^k2^{2k}}{(2k+1)!},~~
{\cal C}_0'^{(k)}=1.~~~~~
\end{eqnarray}
The boundary operator $\tilde J(\tilde z)$ on the sliver frame with
dimension $1$
is related to the one on the unit disk $w=e^{2i\tilde z}$, which we
denote by $J_w(w)$, as $\tilde J(\tilde z)=|2 i e^{2i \tilde
z}|J_w(e^{2i\tilde z})
$. Using $J_w$ and inserting $1=U_1^{-1}U_1$ in front of the
first term, $\Psi_{\lambda_{\rm m}}^{\rm S/KORZ}$ (\ref{eq:psiSKORZv2}) 
can be rewritten as
\begin{eqnarray}
 \Psi_{\lambda_{\rm m}}^{\rm S/KORZ}
&=&-2U_1^{-1}\sum_{k=0}^{\infty}(-\lambda_{\rm m})^{k+1}\!\!
\int_{\cal D}\!\!d\varphi_1^{(k)}
\cdots d\varphi_k^{(k)}
J_w(1)
\prod_{l=1}^k
J_w(e^{i\varphi_l^{(k)}})c_1|0\rangle
\nonumber\\
&&+O(K_1,{\cal L}_0-{\cal L}_0^{\dagger},{\cal B}_0-{\cal
 B}_0^{\dagger},c_n+(-1)^nc_{-n}),
\label{eq:SKORZv3}
\end{eqnarray}
where the arguments are given by changing the variables as
\begin{eqnarray}
 \varphi_l^{(k)}&\equiv &\frac{4(\tilde x_l^{(k)}-\tilde
  x_0^{(k)})}{\gamma^{(k)}}
=-2\pi \frac{\sum_{m=1}^lr_m}{1+\sum_{m=1}^kr_m}\,.
\end{eqnarray}
It induces the Jacobian
\begin{eqnarray}
\left|
 \frac{\partial(\varphi_1^{(k)}\!,\cdots ,\varphi_k^{(k)})}{
\partial(\,r_1\,,\cdots,\,r_k~)}
\right|&=&\frac{
(2\pi)^k}{\left(\gamma^{(k)}\right)^{k+1}}
=\frac{
(2\pi)^k}{\left(1+\sum_{l=1}^kr_l\right)^{k+1}}\,,
\end{eqnarray}
which cancels the extra factor of the first term of
(\ref{eq:psiSKORZv2}).
The integration region ${\cal D}$ is determined by
$0\le r_l\le 1$ ($l=1,2,\cdots,k$) or
\begin{eqnarray}
 0\le -\varphi_1^{(k)}\le 2\pi-(-\varphi_k^{(k)}),~~
0\le -\varphi_{l}^{(k)}-(-\varphi_{l-1}^{(k)})\le 
2\pi-(-\varphi_k^{(k)}),~~(l=2,3,\cdots, k)
\nonumber\\
\label{eq:regionD}
\end{eqnarray}
and the volume is computed as
\begin{eqnarray}
\int_{\cal D}d\varphi_1^{(k)}\cdots d\varphi_k^{(k)} 1=
 \int_0^1\!dr_1\cdots \int_0^1\!dr_k
\frac{(2\pi)^k}{(1+\sum_{l=1}^kr_l)^{k+1}}
=\frac{(2\pi)^k}{(k+1)!}\,.
\label{eq:volumecalD}
\end{eqnarray}
We note that the nonsingular current $J_w$s can be exchanged
without singular behavior.
Therefore, by noting (\ref{eq:levelmatching})
and using 
$U_1^{-1}=(\int_0^{2\pi}\frac{d\theta}{2\pi}e^{\frac{\theta}{4}K_1}
+\int_0^{2\pi}\frac{d\theta}{2\pi}(1-e^{\frac{\theta}{4}K_1})
)U_1^{-1}
=U_1^{-1}\int_0^{2\pi}\frac{d\theta}{2\pi}e^{\frac{\theta}{2}K_1}+O(K_1)$,
\sloppy
the integration in (\ref{eq:SKORZv3}) can be rewritten as that on the
unit circle:
\begin{eqnarray}
 \Psi_{\lambda_{\rm m}}^{\rm S/KORZ}
&=&
-\frac{1}{\pi}U_1^{-1}\sum_{k=0}^{\infty}(-\lambda_{\rm
m})^{k+1}\!
\int_0^{2\pi}\!\!d\theta \!
\int_{\cal D}\!\!d\varphi_1^{(k)}
\cdots d\varphi_k^{(k)}
\nonumber\\
&&~~~~~~~~~~~~~~~~\times 
J_w(e^{i\theta})
\prod_{l=1}^k
J_w(e^{i(\theta+\varphi_l^{(k)})})c_1|0\rangle
\nonumber\\
&&+O(K_1,{\cal L}_0-{\cal L}_0^{\dagger},{\cal B}_0-{\cal
 B}_0^{\dagger},c_n+(-1)^nc_{-n})\,,\nonumber\\
&=&-\frac{1}{\pi}U_1^{-1}
\left(e^{-\lambda_{\rm m}\int_0^{2\pi} d\theta J_w(e^{i\theta})}-1
\right)c_1|0\rangle
\nonumber\\
&&+O(K_1,{\cal L}_0-{\cal L}_0^{\dagger},{\cal B}_0-{\cal
 B}_0^{\dagger},c_n+(-1)^nc_{-n})\,.
\label{eq:SKORZv4}
\end{eqnarray}
In the last equality, we note that any $k+1$ points
specified by coordinates $\theta_l(\in {\mathbb R}\,{\rm mod}\,2\pi)$
$(l=0,1,\cdots,k)$ can be chosen, such as 
$(\theta_0=\theta,\theta_1=\theta+\varphi_1^{(k)},
\cdots,\theta_k=\theta+\varphi_k^{(k)})$,
which satisfy (\ref{eq:regionD})  by shifting the origin and
exchanging them for each other appropriately,
and we have used (\ref{eq:volumecalD}) 
for the $O(\lambda_{\rm m}^{k+1})$ term on multiplicity.

In the case of FKP/KO's marginal solution (\ref{eq:FKP/KOv1}), 
we can also rewrite 
$\Psi^{\rm FKP/KO}_{\lambda_{\rm m},L}$ in the same way as above:
\begin{eqnarray}
\Psi^{\rm FKP/KO}_{\lambda_{\rm m},L}&=&
\sum_{n=1}^{\infty}
\lambda_{\rm m}^n\tilde J\Bigl(\frac{\pi(n-1)}{4n}\Bigr)
\left(\frac{-1}{n}\right)^{n-1}\!
\int^{\frac{\pi(n-1)}{4}}_{\frac{\pi(n-3)}{4}}\!\!d\tilde x_1
\int^{\tilde x_1}_{\frac{\pi(n-5)}{4}}\!\!d\tilde x_2\cdots\!
\int^{\tilde x_{n-2}}_{\frac{-\pi(n-1)}{4}}\!\!d\tilde x_{n-1}
\nonumber\\
&&~~~~~~~~~~~
\times \tilde J(\tilde x_1/n)\tilde J(\tilde x_2/n)\cdots 
\tilde J(\tilde x_{n-1}/n)c_1|0\rangle
\nonumber\\
&&+O({\cal L}_0-{\cal L}_0^{\dagger},c_n+(-1)^nc_{-n})
\nonumber\\
&=&
-\frac{1}{\pi}U_1^{-1}
\sum_{n=1}^{\infty}
(-\lambda_{\rm m})^n\int^{2\pi}_0\!\!d\theta_0
\int^{\theta_0}_{\theta_0-\frac{2 \pi}{n}}\!
d\theta_1\int^{\theta_1}_{\theta_0-\frac{4\pi}{n}}\!d\theta_2
\cdots \int^{\theta_{n-2}}_{\theta_0-\frac{2\pi(n-1)}{n}}\!d\theta_{n-1}
\nonumber\\
&&~~~~~~~~~~~~~~
\times J_w(
e^{i\theta_0})
\prod_{l=1}^{n-1}J_w(e^{i\theta_l})c_1|0\rangle
\nonumber\\
&&+O(K_1,{\cal L}_0-{\cal L}_0^{\dagger},c_n+(-1)^nc_{-n})\,.
\label{eq:FKP/KOv2}
\end{eqnarray}
In this case, noting the volume
\begin{eqnarray}
 \int^{2\pi}_0\!\!d\theta_0
\int^{\theta_0}_{\theta_0-\frac{2 \pi}{n}}\!
d\theta_1\int^{\theta_1}_{\theta_0-\frac{4\pi}{n}}\!d\theta_2
\cdots
\int^{\theta_{n-2}}_{\theta_0-\frac{2\pi(n-1)}{n}}\!d\theta_{n-1}1
&=&\frac{(2\pi)^n}{n!}
\end{eqnarray}
and using the fact\cite{Ellwood} that any $n$ points
specified by coordinates $\theta_l(\in {\mathbb R}\,{\rm mod}\,2\pi)$
$(l=0,1,\cdots,n-1)$ can be chosen, such as 
$\theta_0-\theta_l\le \frac{2\pi}{n}l, \theta_0\ge \theta_1\ge \cdots
\ge \theta_{n-1}$,
by shifting the origin and exchanging them 
for each other appropriately,
(\ref{eq:FKP/KOv2}) can be rewritten as
\begin{eqnarray}
\Psi^{\rm FKP/KO}_{\lambda_{\rm m},L}&=&
-\frac{1}{\pi}U_1^{-1}
\left(e^{-\lambda_{\rm m}\int_0^{2\pi} d\theta J_w(e^{i\theta})}-1
\right)c_1|0\rangle
\nonumber\\
&&+O(K_1,{\cal L}_0-{\cal L}_0^{\dagger},c_n+(-1)^nc_{-n})\,.
\label{eq:FKP/KOv3}
\end{eqnarray}

{}From decompositions of Schnabl/KORZ's (\ref{eq:SKORZv4}) 
and FKP/KO's (\ref{eq:FKP/KOv3}) marginal solutions, the
symmetries of the on-shell closed string state $\langle {\cal I}|V(i)$:
(\ref{eq:symm_Kn}), (\ref{eq:symm_bn}), (\ref{eq:symm_cn}),
and $\langle {\cal I}|V(i)|\varphi\rangle=\langle V(i) f_{\cal I}\circ
\varphi\rangle$, where the conformal map $f_{\cal I}(z)=2z/(1-z^2)$ 
corresponds to $U_1$, we have 
\begin{eqnarray}
 {\cal O}_V(\Psi^{\rm S/KORZ}_{\lambda_{\rm m}})
&=&{\cal O}_V(\Psi^{\rm FKP/KO}_{\lambda_{\rm m},L})
\nonumber\\
&=&-\frac{1}{2\pi i}\left\langle 
c^w\bar c^{w} V_{w\bar w,{\rm m}}(0,0)c^w(1)\left(e^{-\lambda_{\rm
					     m}\int_0^{2\pi} 
d\theta J_w(e^{i\theta})}-1
\right)\right\rangle_{\rm disk}.~~~~
\label{eq:GIO_m_formula}
\end{eqnarray}
The second equality was already shown in Ref.~\citen{Ellwood}.
The first equality means that 
 for the same parameter $\lambda_{\rm m}$ and nonsingular current $J$,
Schnabl/KORZ's marginal solution and 
FKP/KO's one give the same value for the gauge invariant overlap ${\cal
O}_V(\Psi)$.
This is consistent with the expectation that 
$\Psi^{\rm S/KORZ}_{\lambda_{\rm m}}$ and $\Psi^{\rm
FKP/KO}_{\lambda_{\rm m},L}$ 
are gauge equivalent.
The above value (\ref{eq:GIO_m_formula})
is related to the closed string one-point
function\cite{Ellwood}, such as
${\cal A}_{\lambda_{\rm m}}(V)-{\cal A}_{\lambda_{\rm m}=0}(V)$,
where ${\cal A}_{\lambda_{\rm m}}(V)$ is the disk amplitude
for a closed string vertex $V$ with the boundary condition 
deformed by $\lambda_{\rm m}J$.

\section{Comments on rolling tachyon solutions
\label{sec:rolling}
}

Let us consider the gauge invariant overlap ${\cal
O}_{V_{\zeta}}(\Psi)$
with the zero momentum graviton $V_{\rm m}=\zeta_{\mu\nu}\partial
X^{\mu}\bar \partial X^{\nu}$ for 
Hellerman-Schnabl's solution\cite{HS}, which we denote by
$\Psi^{\rm HS}_{\lambda_{\rm m}}$.
$\Psi^{\rm HS}_{\lambda_{\rm m}}$ is given by
Schnabl/KORZ's solution $\Psi^{\rm S/KORZ}_{\lambda_{\rm m}}$ 
with the lightlike rolling tachyon operator $J=e^{\beta X^+}$,
($\beta\equiv 1/(\alpha' V^+)$), on the linear dilaton background
$\Phi(x)=V_{\mu}x^{\mu},(V^+>0,26=D+6\alpha'V_{\mu}V^{\mu},
\mu=0,1,\cdots,D-1)$.
On the linear dilaton background, the matter Virasoro operator is
deformed by $V^{\mu}$ as
$L_n^{(\rm m)}=\frac{1}{2}\sum_k\!:\!\!\alpha_{k,\mu}\alpha_{n-k}^{\mu}\!\!:
+i\sqrt{\frac{\alpha'}{2}}(n+1)V_{\mu}\alpha_n^{\mu}$.
Therefore, polarization $\zeta_{\mu\nu}$ 
for the on-shell closed string state should satisfy the
transversality condition
$\zeta_{\mu\nu}V^{\nu}=V^{\mu}\zeta_{\mu\nu}=0$.
Applying the formula for gauge invariant overlap (\ref{eq:GIO_m_formula}),
we have
\begin{eqnarray}
{\cal O}_{V_{\zeta}}(\Psi^{\rm HS}_{\lambda_{\rm m}})
&=&\frac{1}{2\pi i}
\zeta_{\mu\nu}
\left\langle \partial_wX^{\mu}\bar\partial_{\bar w} X^{\nu}
\left(
e^{-\lambda_{\rm m} \int_0^{2\pi} d\theta J_w(e^{i\theta})}-1
\right)\right\rangle_{\rm disk}^{\rm mat}
\nonumber\\
&=&\int\! d^Dx\,
\frac{1}{2\pi i}
\zeta_{\mu\nu}({\cal A}^{\mu\nu}(x)|_{\lambda=2\pi \lambda_{\rm m}}
-{\cal A}^{\mu\nu}(x)|_{\lambda=0})\,.
\end{eqnarray}
Here, we have used 
\begin{eqnarray}
 {\cal A}^{\mu\nu}(x)&\equiv &\langle 
:\!\partial_wX^{\mu}\bar\partial_{\bar w} X^{\nu}\!:\!(0,0)\, e^{-
\frac{\lambda}{2\pi}\int_0^{2\pi} d\theta J_w(e^{i\theta})}
\rangle_{{\rm disk},x}^{\rm mat}\,,
\end{eqnarray}
which is a CFT correlator in the linear dilaton background on a disk
with a fixed zero mode such as
$x^{\mu}=\frac{1}{2\pi}\int_0^{2\pi} d\theta X^{\mu}(e^{i\theta})$.
Substituting a concrete expression for ${\cal A}^{\mu\nu}(x)$,
which was explicitly computed in \S 5 in Ref.~\citen{HS},
the gauge invariant overlap is evaluated as
\begin{eqnarray}
&&{\cal O}_{V_{\zeta}}(\Psi^{\rm HS}_{\lambda_{\rm m}})
=
\frac{\alpha'}{4\pi i}\!
\int\! d^D\!xe^{-V\!\cdot x}\!\biggl(\!
\zeta_{\mu\nu}\eta^{\mu\nu}(e^{-2\pi \lambda_{\rm m} e^{\beta x^+}}\!\!
-1)
-4\pi \beta^2\alpha' \lambda_{\rm m}
\zeta_{--}e^{\beta x^+-2\pi\lambda_{\rm m}
 e^{\beta x^+}}\!\!\biggr).
\nonumber\\
\label{eq:GIO_HS}
\end{eqnarray}
On the other hand, the gauge invariant overlap ${\cal
O}_{V_{\zeta}}(\Psi)$ for Schnabl's solution for tachyon
condensation $\Psi_{\lambda=1}^{\rm S}$ can be evaluated
using the result in Eq.~(3.28) in Ref.~\citen{KKT1} with the normalization
\begin{eqnarray}
&&C_{V_{\zeta}}=
(2\pi)^D\delta^D\!(iV)\zeta_{\mu\nu}\eta^{\mu\nu}\frac{-\alpha'}{2}
=-\frac{\alpha'}{2}
\zeta_{\mu\nu}\eta^{\mu\nu}\!\int\! d^Dxe^{-V\!\cdot x}~~~~~~
\end{eqnarray}
in this case. Namely, we have
\begin{eqnarray}
 {\cal O}_{V_{\zeta}}(\Psi_{\lambda=1}^{\rm S})&=&
-\frac{\alpha'}{4\pi i}
\zeta_{\mu\nu}\eta^{\mu\nu}\!\int\! d^Dxe^{-V\!\cdot x}\,.
\end{eqnarray}
Comparing the above expression 
and (\ref{eq:GIO_HS}), we obtain the relation
\begin{eqnarray}
\lim_{\lambda_{\rm m}\to +\infty}{\cal O}_{V_{\zeta}}(\Psi^{\rm
 HS}_{\lambda_{\rm m}})
&=&{\cal O}_{V_{\zeta}}(\Psi^{\rm S}_{\lambda=1})\,,
\end{eqnarray}
at least formally.
The result is consistent with the limit of the string field itself, 
$
\lim_{x^+\to +\infty}\Psi^{\rm
 HS}_{\lambda_{\rm m}}
=\lim_{\lambda_{\rm m}\to +\infty}\Psi^{\rm
 HS}_{\lambda_{\rm m}}
=
\Psi^{\rm S}_{\lambda=1}$,
which was proved in Ref.~\citen{HS} in terms of the
${\cal L}_0$ basis.\\

Next, let us consider the ordinary timelike 
rolling tachyon solution, namely, 
Schnabl/KORZ's solution $\Psi^{\rm S/KORZ}_{\lambda_{\rm m}}$ 
with the timelike rolling tachyon operator $J=e^{X^0}$
on the flat background.
It is known\cite{Schnabl:2007az, Kiermaier:2007ba}
 that the tachyon component
given by the coefficient function for $c_1|0\rangle$
in $\Psi^{\rm S/KORZ}_{\lambda_{\rm m}}$  wildly
oscillates for $x^0\to +\infty$
(or $\lambda_{\rm m}\to +\infty$) numerically.
(See also Refs.~\citen{Ellwood:2007xr, Jokela:2007dq} and \citen{HS}.)
This seems to imply  $\lim_{\lambda_{\rm m}\to +\infty}\Psi^{\rm
 S/KORZ}_{\lambda_{\rm m}}
\ne \Psi^{\rm S}_{\lambda=1}$ for  $J=e^{X^0}$.
On the other hand,
one can formally 
evaluate the gauge invariant overlap
with the zero momentum graviton $V_{\rm m}=\zeta_{\mu\nu}2 \partial
X^{\mu}\bar \partial X^{\nu}$ in the same way as in
the above lightlike case using 
the formula (\ref{eq:GIO_m_formula}) and ${\cal A}^{\mu\nu}(x)$
computed in Ref.~\citen{LNT}:
\begin{eqnarray}
 &&{\cal O}_{V_{\zeta}}(\Psi_{\lambda_{\rm m}}^{\rm S/KORZ})
=\frac{1}{2\pi i}\!\int\! d^d\!x\,
\zeta_{\mu\nu}\eta^{\mu\nu}(f(x^0)-1),~~~~f(x^0)\equiv \frac{1}{1+2\pi
\lambda_{\rm m} e^{x^0}}\,.
\end{eqnarray}
If one adopts the limit $\lambda_{\rm m}\to +\infty$ in the
integrand naively, it seems to converge to 
the value for 
Schnabl's solution for tachyon condensation:
${\cal O}_{V_{\zeta}}(\Psi^{\rm S}_{\lambda=1})=-\frac{1}{2\pi
i}\zeta_{\mu\nu}\eta^{\mu\nu}\int d^d\!x$.
However, the limit for the flat space may be too naive because 
$\lambda_{\rm m}$-dependence should be absorbed by shifting the origin
of $x^0$ as an integration value of ${\cal
O}_{V_{\zeta}}(\Psi_{\lambda_{\rm m}}^{\rm S/KORZ})$.
It is desired to define and evaluate {\it local} gauge invariant quantities
in string field theory in order to investigate such a limit.


\section*{Acknowledgments}

The author would like to thank Koji Hashimoto, Hiroyuki Hata,
Teruhiko Kawano, Michael Kiermaier, Yuji Okawa and Tomohiko Takahashi
for helpful discussions and the Yukawa Institute for Theoretical Physics
at Kyoto University for providing a stimulating atmosphere.
Discussions during the YITP workshop YITP-W-08-04 on ``Development of
Quantum Field Theory and String Theory'' were useful in completing this
work.
The work was supported in part by the Special Postdoctoral Researchers
Program at RIKEN and a Grant-in-Aid for Young
Scientists (\#19740155) from the Ministry of Education,
Culture, Sports, Science and Technology of Japan.

%

\appendix

\section{On Schnabl's Solution for Tachyon Condensation
\label{sec:Schn_tach}
}

Schnabl's solution
 for tachyon condensation\cite{Schnabl_tach} is 
similar to Schnabl/KORZ's
marginal solutions but simpler than them.
Hence, it is instructive to investigate the decomposition of
 the solution for tachyon condensation
in order to simplify the computation of gauge invariant overlaps.

Schnabl's solution $\Psi_{\lambda}^{\rm S}$
 with parameter $\lambda$ is
 given by 
\begin{eqnarray}
\Psi_{\lambda}^{\rm S}&=&\frac{\lambda\partial_r}{\lambda
 e^{\partial_r}-1}\psi_r|_{r=0}=\sum_{n=0}^{\infty}\frac{f_n(\lambda)}{n!}
\partial_r^n\psi_r|_{r=0},
\label{eq:Schn_tach}
\\
\psi_r&=&\frac{2}{\pi}\hat{U}_{r+2}
\Biggl[-\frac{1}{\pi}({\cal B}_0+{\cal B}_0^{\dagger})\,
\tilde c\!\left(
\frac{\pi r}{4}\right)
\tilde c\!\left(
\frac{-\pi r}{4}\right)+\frac{1}{2}\left(\tilde c\!\left(
\frac{\pi r}{4}\right)+\tilde c\!\left(
\frac{-\pi r}{4}\right)\right)
\Biggr]|0\rangle.~~~
\label{eq:psirv1}
\end{eqnarray}
The string field $\psi_r$ can be rewritten as
\begin{eqnarray}
 \psi_r&=&\frac{2}{\pi}(1+r)^{\frac{{\cal L}_0-{\cal L}_0^{\dagger}}{2}}
\Biggl[\frac{1}{2}\left(\tilde c\!\left(
\frac{\pi r}{4(1+r)}\right)+\tilde c\!\left(
\frac{-\pi r}{4(1+r)}\right)\right)\nonumber\\
&&
+\frac{1+r}{\pi}({\cal B}_0-{\cal B}_0^{\dagger})\,
\tilde c\!\left(
\frac{\pi r}{4(1+r)}\right)
\tilde c\!\left(
\frac{-\pi r}{4(1+r)}\right)
\Biggr]|0\rangle
\label{eq:psirv2}
\end{eqnarray}
using
\begin{eqnarray}
&&\hat{U}_{r+2}=e^{-\frac{r}{2}({\cal L}_0+{\cal L}_0^{\dagger})}
=(1+r)^{\frac{{\cal L}_0-{\cal L}_0^{\dagger}}{2}}
(1+r)^{-{\cal L}_0}\,,
\label{eq:hatU_identity}
\\
&&\{{\cal B}_0,\tilde c(\tilde z)\}=\tilde z\,,
~~~~~[{\cal L}_0,{\cal B}_0+{\cal B}_0^{\dagger}]
={\cal B}_0+{\cal B}_0^{\dagger}\,.
\label{eq:calB_identity}
\end{eqnarray}
Furthermore, noting the anticommutation relation
$\{b_p,\tilde c(x)\}=(1/2)\sin 2 x
(\tan x)^p$, we have 
\begin{eqnarray}
&&\frac{1}{2}(\tilde c(x)+\tilde c(-x))|0\rangle=
c_1|0\rangle+\cos^2\! x\sum_{k=1}^{\infty}(\tan x)^{2k}(c_{1-2k}-c_{2k-1}
)|0\rangle,
\label{eq:ctil+ctil}\\
&&\tilde c(x)\tilde c(-x)|0\rangle
\nonumber\\
&&=\biggl(
c_0+\sum_{l=1}^{\infty}(\tan x)^{2l}(c_{-2l}+c_{2l})\biggr)\sin x
\biggl(
c_1+\cos^2\! x\sum_{k=1}^{\infty}(\tan x)^{2k}(c_{1-2k}-c_{2k-1})
\biggr)|0\rangle,\nonumber\\
\label{eq:ctilctil}
\end{eqnarray}
which imply that $\psi_r$ (\ref{eq:psirv2}) can be rewritten as
\begin{eqnarray}
 \psi_r&=&\frac{2}{\pi}c_1|0\rangle
+O(
{\cal L}_0-{\cal L}_0^{\dagger},{\cal B}_0-{\cal B}_0^{\dagger},
c_{k}+(-1)^kc_{-k}
).
\end{eqnarray}
Here, $O(
{\cal L}_0-{\cal L}_0^{\dagger},{\cal B}_0-{\cal B}_0^{\dagger},
c_{k}+(-1)^kc_{-k}
)$ denotes some linear combination of terms comprising
${\cal L}_0-{\cal L}_0^{\dagger}$, ${\cal B}_0-{\cal
B}_0^{\dagger}$, and $c_{k}+(-1)^kc_{-k}$,
where at least one of them is multiplied on
the conformal vacuum $|0\rangle$.
The first term $(2/\pi)c_1|0\rangle$ does not depend on $r$.
Using this fact and (\ref{eq:Schn_tach}), we have
\begin{eqnarray}
 \Psi_{\lambda}^{\rm S}&=&\left\{
\begin{array}[tb]{lc}
\frac{2}{\pi}c_1|0\rangle
+O(
{\cal L}_0-{\cal L}_0^{\dagger},{\cal B}_0-{\cal B}_0^{\dagger},
c_{k}+(-1)^kc_{-k})\,, &(\lambda=1)\\
O(
{\cal L}_0-{\cal L}_0^{\dagger},{\cal B}_0-{\cal B}_0^{\dagger},
c_{k}+(-1)^kc_{-k})\,. &(\lambda\ne 1)
\end{array}
\right.
\end{eqnarray}
Because ${\cal L}_0-{\cal L}_0^{\dagger}$
and ${\cal B}_0-{\cal B}_0^{\dagger}$
are linear combinations of $K_n$ and $b_n-(-1)^nb_{-n}$,
respectively, the terms in
$O(
{\cal L}_0-{\cal L}_0^{\dagger},{\cal B}_0-{\cal B}_0^{\dagger},
c_{k}+(-1)^kc_{-k}
)$
give no contribution to the gauge invariant overlaps
thanks to symmetries (\ref{eq:symm_Kn}), (\ref{eq:symm_bn})
and (\ref{eq:symm_cn}) of on-shell closed string states
$\langle {\cal I}|V(i)$.
The first term $\psi_0=\frac{2}{\pi}c_1|0\rangle$ of 
$ \Psi_{\lambda=1}^{\rm S}$ only contributes to the
gauge invariant overlaps, which is consistent with the result in 
Refs.~\citen{Ellwood} and \citen{KKT1},
and it reproduces the ordinary boundary state by contracting the
Shapiro-Thorn vertex with
projection ${\cal P}b_0^-$ in the closed string sector\cite{KKT2}.

\section{On FKP/KO's Marginal Solution
\label{sec:FKP-KO}
}

Here, we review FKP/KO's marginal solution with
 nonsingular current $J$.
Let us construct a solution with parameter $\lambda_{\rm m}$, such as
\begin{eqnarray}
 \Psi=\sum_{n=1}^{\infty}\lambda_{\rm m}^n\psi_n,~~~~
\psi_m=Q_{\rm B}\phi_m+\sum_{k=1}^{m-1}\psi_{k}*\phi_{m-k},
~(m\ge 2);~\psi_1=Q_{\rm B}\phi_1.
\label{eq:dfn_psi_m}
\end{eqnarray}
In fact, using (\ref{eq:dfn_psi_m}) for any $\phi_m$ with ghost number
$0$,
we can check the equation of motion $Q_{\rm B}\Psi+\Psi*\Psi=0$
formally order by order in $\lambda_{\rm m}$:
\begin{eqnarray}
&&Q_{\rm B}\psi_1=0,~~~~
Q_{\rm B}\psi_n+\sum_{k=1}^{n-1}\psi_{n-k}*\psi_k=0.~~~(n\ge 2)
\end{eqnarray}
By choosing $\phi_n$, such as
\begin{eqnarray}
 \phi_n=\frac{(-1)^{n-1}}{n!}X^n(0)|0\rangle*\hat{U}_n|0\rangle=
\frac{(-1)^{n-1}}{n!}\hat{U}_{n+1}
\tilde X^n\Bigl(\frac{\pi}{4}(n-1)\Bigr)|0\rangle,~~(n\ge 1)~~
\label{eq:FKP_phi_n}
\end{eqnarray}
with $X\equiv \zeta_{\mu}X^{\mu},(\zeta_{\mu}\zeta^{\mu}=0)$, one can show
that the obtained solution $\Psi$ is independent of
the zero mode of $X(z)$ \cite{Fuchs:2007yy}.
Noting the BRST transformation $[Q_{\rm B},X(z)]=cJ(z)$ with 
$J\equiv \partial X$, which is a primary field with dimension $1$,
we can obtain $\psi_n$ such as
$\psi_1=c J(0)|0\rangle,\psi_2=-\hat{U}_3\tilde c 
\tilde J(\frac{\pi}{4})\int^{\frac{\pi}{4}}_{-\frac{\pi}{4}}
d\tilde x \tilde J(\tilde x)|0\rangle$ and 
\begin{eqnarray}
\psi_n&=&-\frac{(-1)^{n-2}}{(n-1)!}c J
X^{n-1}(0)|0\rangle*\hat{U}_n|0\rangle
+c J(0)|0\rangle*\phi_{n-1} +\sum_{k=2}^{n-1}\psi_k*\phi_{n-k},
\label{eq:psindef2}
\end{eqnarray}
for $n\ge 3$. Taking the ansatz for $\psi_n$ without $X$ itself:
\begin{eqnarray}
 \psi_n&=&\hat{U}_{n+1}\tilde c\tilde J\Bigl(\frac{\pi}{4}(n-1)\Bigr)
X_{1,2}^{(n)}f_{n-2}(X_{1,2}^{(n)},X_{1,3}^{(n)},
\cdots,X_{1,n}^{(n)})|0\rangle,\\
X_{i,j}^{(n)}&\equiv &\int^{\frac{\pi}{4}(n-2 i+1)}_{\frac{\pi}{4}(n-2
 j+1)}d\tilde x \tilde J(\tilde x), 
\end{eqnarray}
where $f_{n-2}(x_1,\cdots,x_{n-1})$ is a homogeneous polynomial of
degree $n-2$ with respect to $x_i$, and using the star product formula
developed in Ref.~\citen{Schnabl_tach}, we find the recurrence equation
\begin{eqnarray}
f_{n-2}(x_1,x_2,
\cdots,x_{n-1})
&=&-\frac{(x_1)^{n-2}}{(n-1)!}
-\sum_{k=0}^{n-3}
f_k(x_1,x_2,\cdots,x_{k+1})
\frac{(x_{k+2})^{n-k-2}}{(n-k-2)!}~~~
\label{eq:recurrence2}
\end{eqnarray}
$(n\ge 3)$ with $f_0\equiv -1$
from (\ref{eq:psindef2}) and (\ref{eq:FKP_phi_n}).
This equation can be solved as
\begin{eqnarray}
&&x_1f_{n-2}(x_1,x_2,\cdots, x_{n-1})
\nonumber\\
&&=-\frac{(x_1)^{n-1}}{(n-1)!}+\sum_{s=1}^{n-2}(-1)^{s-1}
\sum_{k_1=1}^{n-2-(s-1)}\cdots
\sum_{k_p=1}^{n-2-(s-p)-\sum_{l=1}^{p-1}k_l}
\cdots
\sum_{k_s=1}^{n-2-\sum_{l=1}^{s-1}k_l}\Biggl[
\nonumber\\
&&~~~~~~~~~~~~~~~~
\frac{(x_1)^{k_1}}{(k_1)!}\!
\left(\prod_{q=1}^{s-1}
\frac{(x_{\sum_{m=1}^qk_m+1})^{k_{q+1}}}{(k_{q+1})!}\right)
\!\frac{(x_{\sum_{m=1}^{s}k_m+1})^{n-1-\sum_{m=1}^sk_m}}
{(n-1-\sum_{m=1}^sk_m)!}\Biggr].
\end{eqnarray}
Using the above result, we have
\begin{eqnarray}
&&X_{1,2}^{(n)}f_{n-2}(X_{1,2}^{(n)},X_{1,3}^{(n)},
\cdots,X_{1,n}^{(n)})
\nonumber\\
&&=(-1)^{n-1}\!\!\!
\int^{\frac{\pi}{4}(n-1)}_{\frac{\pi}{4}(n-3)}\!\!\!d\tilde x_1
\int^{\tilde x_1}_{\frac{\pi}{4}(n-5)\!}\!\!\!d\tilde x_2
\int^{\tilde x_2}_{\frac{\pi}{4}(n-7)\!}\!\!\!d\tilde x_3
\cdots\!\!
\int^{\tilde x_{n-2}}_{-\frac{\pi}{4}(n-1)\!}\!\!d\tilde x_{n-1}
\tilde J(\tilde x_1)\tilde J(\tilde x_2)
\cdots \tilde J(\tilde x_{n-1}),
\nonumber\\
\end{eqnarray}
which corresponds to the expression given in Ref.~\citen{Kiermaier:2007vu}.
If we use this formula or (\ref{eq:FKP/KOv1}) 
with any matter primary field $J$, which has dimension $1$ 
and nonsingular OPE,
we can check that the obtained string field
$\Psi^{{\rm FKP/KO}}_{\lambda_{\rm m},L}$
satisfies the equation of motion.

\section{Relation to the Shapiro-Thorn Vertex
\label{sec:S-T}
}

It is convenient to use the Shapiro-Thorn vertex $\langle
\hat{\gamma}(1_{\rm c},2)|$
to find formulas related to the gauge invariant overlaps,
because they can be expressed using
 $\langle
\hat{\gamma}(1_{\rm c},2)|$, as in (\ref{eq:GIO_defn}).
On the closed and open string sides, 
$\langle\hat{\gamma}(1_{\rm c},2)|$ is
 specified by maps $h_1(w_1)=-i(w_1-1)/(w_1+1)$
and $h_2(w_2)=(w_2-1/w_2)/2$, respectively.
(See Appendix B in Ref.~\citen{KKT1} for details.)
Using these maps or explicit formulas for Neumann coefficients, we can
derive
\begin{eqnarray}
&&\langle\hat{\gamma}(1_{\rm c},2)|\left(K_n^{(2)}-
(-1)^{\frac{n}{2}}\frac{n}{4}c\,\delta_{n:{\rm even}}\right)
\nonumber\\
&&=
\langle\hat{\gamma}(1_{\rm
c},2)|(-2i^n)\sum_{m\ge
0}(-1)^m(\eta^n_{2m+1}-
\eta^n_{2m-1})(L^{(1)}_m+(-1)^n\bar L^{(1)}_m),
\label{eq:ST-Kn_id}
\\
&&\langle\hat{\gamma}(1_{\rm c},2)|(b_n^{(2)}-(-1)^nb_{-n}^{(2)})
\nonumber\\
&&=\langle\hat{\gamma}(1_{\rm c},2)|
(-2i^n)\sum_{m\ge
0}(-1)^m(\eta^n_{2m+1}-
\eta^n_{2m-1})(b^{(1)}_m+(-1)^n\bar b^{(1)}_m),\\
&&\langle\hat{\gamma}(1_{\rm c},2)|(c_m^{(2)}+(-1)^mc_{-m}^{(2)})
\nonumber\\
&&=\langle\hat{\gamma}(1_{\rm c},2)|
\frac{-i^m}{4}\sum_{n\ge
1}(-1)^n(\eta^{2n}_{m+1}-
\eta^{2n}_{m-1}+\delta_{m,1})
(c^{(1)}_n+(-1)^m\bar c^{(1)}_n),
\end{eqnarray}
where $\delta_{n:{\rm even}}=1(0)$  for $n$: even (odd) and
$c$ is the central charge for the Virasoro algebra in the first line.
$\eta^k_n$ is defined by the generating function 
$\left(\frac{1+x}{1-x}\right)^k=\sum_{n=0}^{\infty}\eta^k_nx^n$.
By contracting the above with the closed string state $|V_c\rangle_{1_{\rm
c}}=c_1\bar c_1V_{\rm m}(0,0)|0\rangle_{1_{\rm c}}$ where
$V_{\rm m}(z,\bar z)$ is a matter primary field with 
 dimension $(1,1)$, we obtain formulas (\ref{eq:symm_Kn}), 
(\ref{eq:symm_bn}) and (\ref{eq:symm_cn}).

We note that the level-matching projection
${\cal P}=\int_0^{2\pi}\frac{d\theta}{2\pi}e^{-i\theta(L_0-\bar{L}_0)}$
for closed string states corresponds to
$\int_0^{2\pi}\frac{d\theta}{2\pi}e^{\frac{\theta}{4}K_1}$
for the open string side on the Shapiro-Thorn vertex because of the identity
\begin{eqnarray}
 \langle\hat{\gamma}(1_{\rm c},2)|(L_0^{(1)}-\bar{L}_0^{(1)})
=\langle\hat{\gamma}(1_{\rm c},2)|\frac{i}{4}K_1^{(2)},
\label{eq:levelmatching}
\end{eqnarray}
which follows from (\ref{eq:ST-Kn_id}).

\end{document}